\documentclass[a4paper,11pt]{article}
\usepackage{jcappub}

\usepackage{caption}
\usepackage{float}
\usepackage{mathtools}

\usepackage{hyperref}

\newcommand{\av}[1]{\langle{#1}\rangle}

\title{Cosmological backreaction within the Szekeres model and emergence of spatial curvature}

\author{Krzysztof Bolejko}
\affiliation{Sydney Institute for Astronomy, School of Physics, A28, The University of Sydney, NSW, 2006, Australia}
\emailAdd{krzysztof.bolejko@sydney.edu.au}

\abstract{This paper discusses the phenomenon of backreaction within the Szekeres model.
Cosmological backreaction describes how the mean global evolution of the Universe deviates from the Friedmannian evolution. The analysis is based on models of a single cosmological environment and the global ensemble of the Szekeres models (of the Swiss-Cheese-type and Styrofoam-type). The obtained results show that non-linear growth of cosmic structures is associated with the growth of the spatial curvature $\Omega_{\cal R}$ (in the FLRW limit $\Omega_{\cal R} \to \Omega_k$). If averaged over global scales the result depends on the assumed global model of the Universe.
Within the Swiss-Cheese model, which does have a fixed background, the volume average follows the evolution of the background, and the global spatial curvature averages out to zero (the background model is the $\Lambda$CDM model, which is spatially flat). In the Styrofoam-type model, which does not have a fixed background, the mean evolution deviates from the spatially flat $\Lambda$CDM model, and the mean spatial curvature evolves from $\Omega_{\cal R} =0 $ at the CMB to $\Omega_{\cal R} \sim 0.1$ at $z =0$. If the Styrofoam-type model correctly captures evolutionary features of the real Universe then one should expect that in our Universe, the spatial curvature should build up (local growth of cosmic structures) and its mean global average should deviate from zero (backreaction).
As a result, this paper predicts that the low-redshift Universe should not be spatially flat (i.e. $\Omega_k \ne 0$, even if in the early Universe $\Omega_k = 0$) and therefore when analysing low-$z$ cosmological data one should keep $\Omega_k$ as a free parameter and independent from the CMB constraints.}

\begin{document}
\maketitle
\flushbottom

\section{Introduction}\label{intro}

Backreaction is a process that describes feedback of structure formation on the mean global evolution of the Universe. As long as the evolution of cosmic structures is well within the linear regime, the Universe should  successfully be approximated by perturbations around the FLRW model and its global evolution should follow the Friedmann solution. This part of our Universe's evolution seems to be well understood. 
What is less understood and still debatable among cosmologists is the current epoch of the Universe's evolution with its non-linear growth of cosmic structures. Some cosmologists argue that in the non-linear regime the average evolution of the Universe may deviate from the Friedmannian evolution.

The Friedmannian evolution is at the core of the standard cosmological model. Observational data are mostly interpreted within this framework. Also this framework is embedded within $N$-body simulations. The $N$-body simulations are employed to trace the evolution of our Universe in the non-linear regime. Mostly this is done by employing the Newtonian physics on small scales, and assuming that the global evolution follows the  Friedmann solution and is unaffected by local interactions.
In principle it is possible to construct a Newtonian cosmology based on a system that expands homothetically
and obey the Friedmannian evolution \cite{2014CQGra..31b5003E}. Also the linear perturbations of such a system  are known to be consistent with linear perturbations of the FLRW model \cite{2015CQGra..32e5001E}. However, in the non-linear regime the situation may be different \cite{Eingorn:2012jm,Eingorn:2012dg}, for example one of the recent studies shows that the relativistic effects lead to Yukawa-type interactions between matter particles \cite{2016ApJ...825...84E}.
Studies based on the weak-filed limit (i.e. applying post-Newtonian corrections)
suggest that the mean evolution is well approximated by the Friedmannian evolution \cite{2013PhRvD..88j3527A,2014CQGra..31w4006A,2016JCAP...07..053A}. On the other hand, studies that try to implement backreaction-type effects into the Newtonian Cosmology and $N$-body codes 
find large effects \cite{2016arXiv160100110K,2016arXiv160708797R} --- such an approach has sparked recently a debate, see Ref. \cite{2017arXiv170308809K} and the rebuttal response in Ref. \cite{2017arXiv170400703B}. 

The presence of backreaction is a mathematical consequence of the non-linear structure of equations that govern the evolution of the Universe \cite{2005PhLA..347...38E}. Cosmological backreaction has been debated for the last 30 years \cite{Buchert:2015iva,2016arXiv161208222B}. 
The nature of this debate focuses on the magnitude of backreaction.
On one hand, predictions derived from models based on the FLRW framework and $N$-body simulations are consistent with observational data; this is inductive evidence suggesting that the backreaction could be negligibly small. On the other hand, presence of tensions between various data and experiments could be related to backreaction effects.
Finally, some people argue that the presence of dark energy, which dominates the energy budget of the standard cosmological model, is just an artefact and nothing else as a manifestation of strong backreaction effects.

This paper aims to be another voice in the debate on the backreaction and presents the study of  the backreaction within the Szekeres model, which is an exact inhomogeneous cosmological solution of the Einstein equations. 
The structure of the paper is as follows: Sec. \ref{evolution} continues with the introduction to the phenomenon of backreaction; Sec. \ref{szekeres} introduces the Szekeres model; Secs. \ref{local} and \ref{global} present the results of the analysis of the backreaction phenomenon within the Szekeres model; Sec. \ref{conclusions} concludes the analysis.

\section{Evolution of matter in the Universe and backreaction}\label{evolution}

The energy momentum tensor of a viscous fluid with no energy transfer can be written as

\begin{equation}
T_{ab}= \rho u_au_b+ ph_{ab}+ \pi_{ab},
\end{equation}
where $\rho$ is energy density, $p$ is pressure, $\pi_{ab}$ is the anisotropic stress tensor, and $h_{ab}$ is the spatial part of the metric in 3+1 split

\begin{equation}
h_{ab}= g_{ab} - u_a u_b,
\end{equation}
and $u_a$ is the matter velocity flow, whose gradient can be decomposed as
\[ u_{a;b} = \omega_{a b} + \sigma_{a b}
+ \frac{1}{3} h_{a b} \Theta - A_a u_b, \]
where
$\omega_{a b} = u_{[a;|\sigma|} h^\sigma{}_{ b]}$ is rotation,
$\sigma_{a b} = u_{(a;|\sigma|} h^\sigma{}_{ b)} - \frac{1}{3} h_{a b} \Theta$ is shear,
$\Theta = u^a{}_{;a}$ is expansion, and $ A^a = u^a{}_{;b} u^b$ is acceleration.

Evolution of expansion, shear, and rotation are given by \cite{
2009GReGr..41..581E,2008PhR...465...61T}
\begin{eqnarray}
&& \dot{\Theta}  = -{1\over3}\,\Theta^2- {1\over2}\,(\rho+3p)-
2(\sigma^2-\omega^2) + {\rm D}^aA_a+ A_aA^a+ \Lambda, \label{ffe1} \\
&&  \dot{\sigma}_{\langle ab\rangle}  = -{2\over3}\,\Theta\sigma_{ab}-
\sigma_{c\langle a}\sigma^c{}_{b\rangle}- \omega_{\langle
a}\omega_{b\rangle} + {\rm D}_{\langle a}A_{b\rangle}+ A_{\langle
a}A_{b\rangle} - E_{ab}+ {1\over2}\,\pi_{ab},\\
&& \dot{\omega}_{\langle a\rangle}  = -{2\over3}\,\Theta\omega_a-
{1\over2}\,{\rm curl} A_a+ \sigma_{ab}\omega^b.
\end{eqnarray}
The equations for density, pressure, and anisotropic stress are \cite{
2009GReGr..41..581E,2008PhR...465...61T}
\begin{eqnarray}
&& \dot{\rho}  = -\Theta(\rho+p)- 
\sigma^{ab}\pi_{ab}, \\    
&&  (\rho+p)A_a  = -{\rm D}_ap-  {\rm D}^b\pi_{ab}- \pi_{ab}A^b.
\end{eqnarray}
Finally, the evolution of the electric ($E_{ab}$) and magnetic ($H_{ab}$) parts of the Weyl curvature are given by \cite{
2009GReGr..41..581E,2008PhR...465...61T}

\begin{eqnarray}
&&  \dot{E}_{\langle ab\rangle}  = -\Theta E_{ab}-
{1\over2}\,(\rho+p)\sigma_{ab}+  {\rm curl} H_{ab}-
{1\over2}\,\dot{\pi}_{ab} -{1\over6}\,\Theta\pi_{ab} \nonumber
\\ && ~~~~~~  +3\sigma_{\langle
a}{}^c\left(E_{b\rangle c}-{1\over6}\,\pi_{b\rangle c}\right) +
\varepsilon_{cd\langle
a}\left[2A^cH_{b\rangle}{}^d-\omega^c\left(E_{b\rangle}{}^d+
{1\over2}\,\pi_{b\rangle}{}^d\right)\right],   \\
&&  \dot{H}_{\langle ab\rangle} = -\Theta H_{ab}- {\rm curl} E_{ab}+
{1\over2}\,{\rm curl} \pi_{ab}  + 3\sigma_{\langle a}{}^cH_{b\rangle c}
-\varepsilon_{cd\langle a}\left(2A^cE_{b\rangle}{}^d+\omega^cH_{b\rangle}{}^d\right). \label{ffe10} 
\end{eqnarray}

The homogeneous and isotropic FLRW models form a special subset of all possible solutions of the above equations. The FLRW solution is characterised by vanishing Weyl curvature, vanishing rotation and shear, zero anisotropic stress and pressure gradients \cite{1997icm..book.....K}

\begin{eqnarray}
{\rm FLRW~universe~~} \left\{ \begin{array}{llllll}
E_{ab} \equiv 0  \nonumber \\
H_{ab}  \equiv 0  \nonumber \\
\omega_{ab}  \equiv 0  \nonumber \\
\sigma_{ab} \equiv 0  \nonumber \\
\pi_{ab} \equiv 0 \nonumber \\
D_a p \equiv 0 \nonumber 
\end{array} \right.
\end{eqnarray}

In such a case all the above given evolution equations reduce only to 2 equations that fully describe the evolution of a spatially homogeneous and isotropic system 
\begin{eqnarray}
\dot{\Theta}  = -{1\over3}\,\Theta^2- {1\over2}\,(\rho+3p) + \Lambda, \label{hfe1} \\
 \dot{\rho}  = -\Theta(\rho+p). \label{hfe2}
\end{eqnarray}

After some algebra, it can be shown that the  above equations are equivalent to the Friedmann equations

\begin{eqnarray}
&& 3 \frac{\ddot{a}}{a} = - 4 \pi G (\rho + 3p)  +  \Lambda,  \label{fes1} \\ 
&&  3 \frac{\dot{a}^2}{a^2} = 8 \pi G  \rho  - 3 \frac{k}{a^2}   + \Lambda  \label{fes2},
\end{eqnarray}
where the relation between the scale factor $a(t)$ and the expansion rate is $\Theta = 3 \dot{a}/a$, and the spatial curvature is ${\cal R} = 6 k /a^2$. 

If the universe is homogeneous, then the average evolution is exactly the same as the evolution of an individual worldline, and is given by the Friedmann equations.
If the universe is inhomogeneous, then the average over all individual worldlines, may deviate from the solution of a uniform universe, and therefore deviate from the Friedmannian evolution.
This is what backreaction describes. When studying backreaction, one focuses on all neglected (in the FLRW case) terms [cf. eq. (\ref{ffe1})--(\ref{ffe10})] and investigates if it is possible 
that  all these terms can affect the mean global evolution of the inhomogeneous system. In other words, if global (mean) evolution of the volume of the Universe (i.e. $\Theta$) and/or matter (i.e. $\rho$) is the same as prescribed by the Friedmann solution [i.e. eqs. (\ref{hfe1}) and (\ref{hfe2}) or equivalently by eqs. (\ref{fes1}) and (\ref{fes2})] or if the contribution from the shear, rotation, and Weyl curvature can affect the expansion rate $\Theta$, and subsequently change its global (mean) evolution compared to the FLRW case.

Because the evolution equations  (\ref{ffe1})--(\ref{ffe10})
 are complicated, we still lack a satisfactory description of backreaction for a real Universe. If we limit the analysis only to irrotational and pressureless fluids (the Szekeres model discussed below can only describe irrotational dust) then the scalar parts of the above equations can be averaged and reduced to \cite{Buchert:1999er}

\begin{eqnarray}
&& 3 \frac{\ddot{a}_{\cal D}}{a_{\cal D}} = - 4 \pi G \av{\rho}_{\cal D} + \Lambda + \mathcal{Q}_{\cal D}, \label{bucherteq1} \\ 
&&  3 \frac{\dot{a}^2_{\cal D}}{a^2_{\cal D}} = 8 \pi G \av{ \rho}_{\cal D}  - \frac{1}{2} \av{ \mathcal{R} }_{\cal D} + \Lambda - \frac{1}{2} \mathcal{Q}_{\cal D}, \label{bucherteq2} \\
&& \big( {\cal Q}_{\cal D} a^6_{\cal D} \dot{\big)} + 
a^4_{\cal D}  \big(   \av{ \mathcal{R} }_{\cal D} a^2_{\cal D} \dot{\big)}  = 0, \label{bucherteq3}
\end{eqnarray}
where the dot $\dot{}\,$ denotes partial time derivative $\dot{} \equiv \partial_t$, $\av{ \mathcal{R} }_{\cal D}$ is an average of the spatial Ricci scalar $\mathcal{R}$,  $\av{\ }_{\cal D}$ is the volume average over the hypersurface of constant time

\[ \av{A}_{\cal D} = \frac{\int_{\cal D} d^3x \sqrt{|h|} A }{\int_{\cal D} d^3x \sqrt{|h|}}, \] 
the scale factor $a_{\cal D}$ is defined as 

\begin{equation}
a_{\cal D} =  \left( \frac{ V_{\cal D} }{V_{{\cal D},i} } \right)^{1/3},
\label{aave}
\end{equation}
where V$_{\cal D}$ is the volume of the domain ${\cal D}$, and $V_{ {\cal D}, i}$ is its initial value.
Finally the function ${\cal Q}_{\cal D}$ is

\begin{equation}
{\cal Q}_{\cal D} = \frac{2}{3} \left( \av{ \Theta^2}_{\cal D} - \av{ \Theta}_{\cal D}^2 \right)
- 2 \av{ \sigma^2}_{\cal D}.
\end{equation}

Equations (\ref{bucherteq2}) can be used to define the Hubble parameter
$H_{\cal D}$

\[ H_{\cal D}^2=  \frac{8 \pi G}{3} \av{ \rho}_{\cal D}  - \frac{1}{6} \av{ \mathcal{R} }_{\cal D} + \frac{1}{3} \Lambda - \frac{1}{6} \mathcal{Q}_{\cal D}, \]
which then can be used to introduce the {\em cosmic quartet} \cite{2008GReGr..40..467B}

\begin{eqnarray}
\Omega_m^{\cal D} &=& \frac{8 \pi G}{3 H_{\cal D}^2 } \av{\rho}_{\cal D}, \nonumber \\
\Omega_\Lambda^{\cal D} &=& \frac{\Lambda}{3 H_{\cal D}^2}, \nonumber \\
\Omega_\mathcal{R}^{\cal D} &=& -\frac{ \av{\mathcal{R}}_{\cal D} } { 6 H_{\cal D}^2 }, \nonumber \\
\Omega_\mathcal{Q}^{\cal D} &=& -\frac{ {\mathcal{Q}}_{\cal D} } { 6 H_{\cal D}^2 }.
\label{cqar}
\end{eqnarray}

It follows from eq. (\ref{bucherteq3}) that iff ${\cal Q}_{\cal D} = 0$ at all times then 
the average spatial curvature reduces to the Friedmannian curvature, i.e. $\av{ \mathcal{R} }_{\cal D} \sim a_{\cal D}^{-2}$. This implies that iff ${\cal Q}_{\cal D}$ vanishes at all times then the Buchert equations reduce to the Friedmann equations, meaning that the 
non-linear effects associated with structure formation cannot affect the average evolution of the Universe.
The question and issue of the ongoing debate on backreaction is related to the amplitude of backreaction. Some studies suggest large backreaction \cite{2006CQGra..23.6379B,2006JCAP...11..003R,2008JCAP...04..026R,2007NJPh....9..377W,2009PhRvD..79h3011L,2009PhRvD..80l3512W,2010PhRvD..82b3523W,2011CQGra..28p4006W,2013JCAP...10..043R,2017A&A...598A.111R}, while some suggest otherwise \cite{Green:2010qy, Green:2014aga, Green:2015bma}.
The debate on whether the backreaction in realistic models of the Universe is strictly zero has already been settled \cite{Buchert:2015iva}. However, it is still debated whether  a small deviation from zero also means almost Friedmannian evolution, or not.
This question is addressed in this paper using the Szekeres solution of the Einstein equations. The investigation is based on a realistic model of the local cosmological environment \cite{2015arXiv151207364B} and the ensemble of Szekeres models that is based on this type of structures.

\section{Szekeres model}\label{szekeres}

The Szekeres model \cite{1975CMaPh..41...55S} is one of the most general inhomogeneous, exact, cosmological solutions of the Einstein equations \cite{1997icm..book.....K,2009suem.book.....B}. 
The metric of the Szekeres model \cite{1975CMaPh..41...55S,1975PhRvD..12.2941S}
in the spherical coordinates is \cite{spheszek}

\begin{eqnarray}
&& {\rm d} s^2 = { {\rm d} t^2} - \frac{1}{\varepsilon-K} \left[ R' + \frac{R}{S} \left( S' \cos \theta + N \sin \theta \right) \right]^2 { {\rm d} r^2 }
- R^2 \left( { {\rm d}{\theta}^2} + \sin^2\theta {{\rm d} {\phi}^2} \right) \nonumber \\ 
&&  + \left( \frac{R}{S} \right)^2 \left\{ 
\left[ S' \sin \theta + N (1 - \cos \theta ) \right]^2 +
 \left[ ( \partial_\phi N )(1-\cos \theta) \right]^2 \right\} { {\rm d} r^2} \nonumber \\ 
&& +2 \left( \frac{R}{S} \right)^2 \left[ S S' \sin \theta + S N (1 -\cos \theta) \right] { {\rm d} r {\rm d} \theta } - 2 \left( \frac{R}{S} \right)^2  S ( \partial_\phi N ) \sin \theta (1-\cos \theta) { {\rm d} r {\rm d} \phi},\label{ds2}
\end{eqnarray}
where ${}' \equiv \partial/\partial r$, $ N(r,\phi) \equiv P' \cos\phi + Q'\sin\phi$,  $\varepsilon = \pm1,0$ and $K = K(r) \leq \varepsilon$,  $S = S(r)$, $P = P(r)$, and $Q = Q(r)$
are arbitrary functions of $r$.

Systems that can be described using the Szekeres solution have no vorticity $\omega_{ab} = 0$, no viscosity $\pi_{ab} =0$, no pressure $p=0$, and no gravitational radiation $H_{ab} =0$. The Szekeres models are of Petrov type D, the shear and electric Weyl tensor can be written as

\[ \sigma_{ab} = \Sigma \, {\rm e}_{ab}, \quad E_{ab} = {\cal W} \, {\rm e}_{ab}, \]
where ${\rm e}_{ab} = h_{ab} - 3 z_a z_b$ where $z^a$ is a space-like unit vector aligned with the Weyl principal tetrad. As a result the fluid equations (\ref{ffe1})--(\ref{ffe10})
reduce only to 4 scalars \cite{2002PhRvD..66h4011H,2009suem.book.....B,2012CQGra..29f5018S} 

\begin{align}
& \rho =  \frac {2 \left(M' - 3 M  E' /  E\right)} {R^2 \left(R' - R E' /  E\right)},\label{rho} \\
& \Theta = \frac{ \dot{R}' + 2 \dot{R}R'/R - 3 \dot{R}  E' /  E}{R' - R E' /  E}, \\
& \Sigma = -\frac{1}{3}  \frac{ \dot{R}' - \dot{R}R'/R}{R' - R E' /  E},  \\
& {\cal W} =  \frac{M}{3R^3} \frac{3R' - R M'/M} { R' - R E' /  E},
\end{align}
where $E'/E = - (S' \cos \theta + N \sin \theta ) /S$. The spatial curvature follows from the ``Hamiltonian'' constraint
\begin{equation}
\frac{1}{6} {\cal R}= \frac{1}{3}  \rho + 6 \Sigma^2  - \frac{1}{9} \Theta^2 + \frac{1}{3} \Lambda, \label{hamcon}
\end{equation}
 and is given by
\begin{equation}
{\cal R} = 2 \frac{K}{R^2} \left( 1 +  \frac{ R K'/K - 2 R E' /  E}{ R' - R E' /  E} \right).
\end{equation}
Thus, the whole evolution of the system, is reduced only to a single equation for the function $R$
\begin{equation}
\dot{R}^2  = -K +\frac{2 M(r)}{R} +\frac 1 3\Lambda R^2. \label{evo} 
\end{equation}

To define the Szekeres model 5 arbitrary functions of radial coordinate $r$ need to be specified.
In this paper, the radial coordinate has been chosen as

\[ r= R_{i}, \]
where $R_i$ is the value of $R$ at the last scattering instant, and the arbitrary functions were chosen to be

\begin{eqnarray}
M(r) &=& \frac{1}{6} 8 \pi G \rho_{i} \left[ 1 + \frac{1}{2} m_0 \left( 1 - \tanh \frac{r-r_0}{2 \Delta r} \right)  \right] r^3, \label{szf1} \\
K(r) &=& \frac{7}{9} 4 \pi G m_0 \rho_i \left( 1 - \tanh \frac{r-r_0}{2 \Delta r} \right)  r^2,\label{szf2}  \\ 
S(r) &=& r^\eta, \label{szf3}  \\
P(r) &=& 0,  \label{szf4} \\
Q(r) &=& 0. \label{szf5} 
\label{szekeresfunctions}
\end{eqnarray}
Thus, the model is prescribed by following parameters:
$r_0$ (size of perturbation), $\Delta r$ (transition zone), $m_0$ (central density contrast of the initial perturbation), $\eta$ (dipole's size), and $\rho_{i}$ (background density at the initial instant). The initial instant is set to be the last scattering instant, so $\rho_i = (1+z_{CMB})^3 \Omega_m 3 H_0^2/(8 \pi G)$, where $z_{CMB}$ is the CMB's redshift $z_{CMB} = 1090$.

\section{Local non-linear evolution and emergence of local spatial curvature within a single cosmological environment}\label{local}

\begin{figure*}
\begin{center}
\includegraphics[scale=0.57]{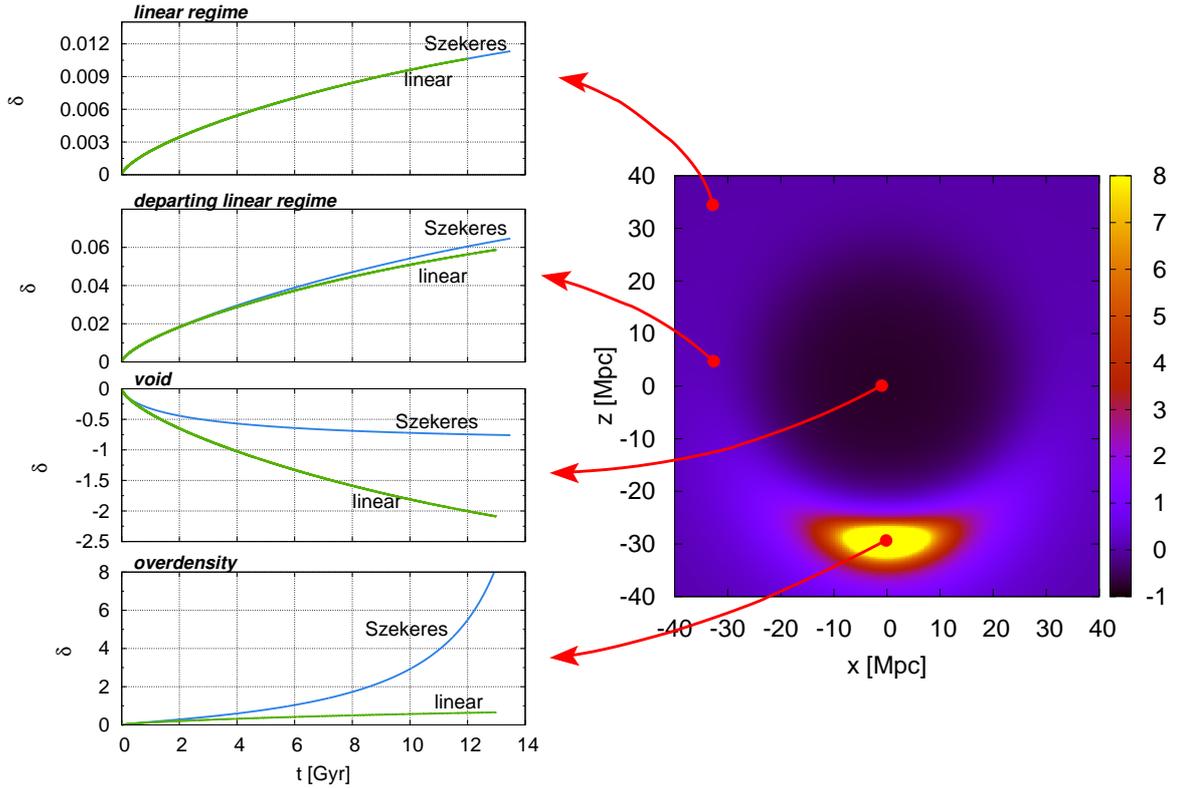}
\end{center}
\caption{Evolution of matter within the studied Szekeres model. The panel on the right shows the present-day density distribution, and the panels on the left show the evolution of the density contrast in places from where the arrows point out. As seen the central cosmic void and large overdensity undergo the non-linear growth.}
\label{fig1}
\end{figure*}

\begin{figure*}
\begin{center}
\includegraphics[scale=0.685]{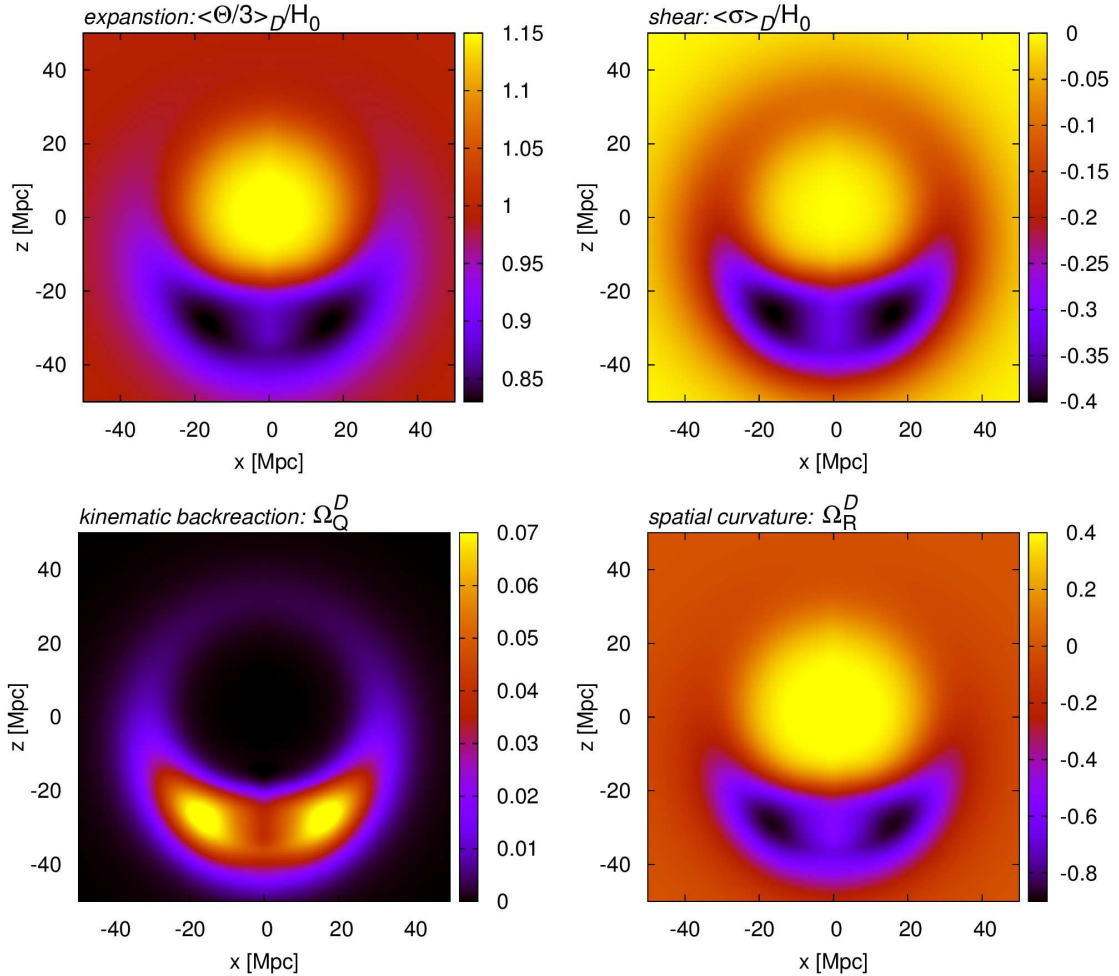}
\end{center}
\caption{The average quantities within the studied Szekeres model, whose density distribution is presented in Fig. \ref{fig1}. Each point shows  a quantity, which is volume averaged (at the current instant) over a domain of radius of 5 Mpc. The upper left panel shows the average expansion rate $\av{\Theta}_{\cal D}/{3 H_0}$; the upper right shows the average shear $\av{\sigma}_{\cal D}/H_0$; the lower left panel shows the kinematic backreaction $\Omega_{\mathcal{Q}}^{\cal D}$ and the lower right panel shows the spatial curvature $\Omega_\mathcal{R}^{\cal D}$.}
\label{fig2}
\end{figure*}

The model of a local cosmological environment is specified by fixing the free parameters to

\begin{align}
& r_0  = 20 {\rm ~Kpc},  \nonumber \\
&\Delta r= \frac{1}{3} r_0,  \nonumber \\
& m_0 = -0.00215,   \nonumber \\
& \eta = 0.52, \nonumber \\
& \rho_i = (1+z_{CMB})^3 \Omega_m 3 H_0^2/(8 \pi G), \nonumber \\
& \Omega_m = 0.315, \nonumber \\
& \Omega_\Lambda = \frac{\Lambda}{3 H_0^2} = 0.685, \nonumber \\
& H_0 = 67.3 {\rm~km} {\rm~s}^{-1} {\rm~Mpc}^{-1}.
\end{align}
The evolution of the system is calculated by solving eq. (\ref{evo}) from the last scattering instant till the present-day instant. The present-day density contrast presented in the right panel of Fig. \ref{fig1} is defined as

\[ \delta = \frac{\rho-\rho_0}{\rho_0}, \]
where $\rho_0$ is the present-day background density, $\rho_0 = 3 H_0^2 \Omega_m/(8 \pi G)$. The evolution of the density contrast at 4 different locations is also presented in Fig. \ref{fig1}. The evolution of the density contrast evaluated within this Szekeres model is compared to the evolution of the  density contrast $\delta_{lin}$
which follows from the linear approximation \cite{1980lssu.book.....P}

\begin{equation}
 \ddot{\delta}_{lin} + 2 \frac{ \dot{a}}{a} \dot{\delta}_{lin} - 4 \pi G \rho \delta_{lin} =0,
\label{lineara}
\end{equation}
where the initial conditions for $\delta_i$ and $\dot{\delta}_i$ have been chosen to be the same as in the Szekeres model at the last scattering instant.

Two upper left panels in Fig. \ref{fig1} show places where the models is quite homogeneous, and where the evolution is well described by the linear approximation. 
The other two panels show places where the inhomogeneity is large and the growth is non-linear.
A more detailed map on local inhomogeneity and averages is presented in Fig. \ref{fig2}. Figure \ref{fig2} presents the volume averages evaluated within spherical domains. Each point in Fig. \ref{fig2} presents the value of the averaging within a domain centred at that point and of radius 5 Mpc (cf. the matter horizon in Ref. \cite{2009MNRAS.398.1527E}). The volume averages were evaluated using the code {\texttt{SzReD}}\footnote{The code {\texttt{SzReD}} is not yet publicly available under a free licence. It is accessible on collaboration basis and anyone interested in using the code is advised to contact the author.}. The code uses the Healpix grid to evaluate integrals around any given point in space and time.
The average expansion rate and shear at the present instant normalised by $H_0$ are presented in the upper panels. The kinematic backreaction $\Omega_{\cal Q}^{\cal D}$ and spatial curvature $\Omega_\mathcal{R}^{\cal D}$ are presented in the lower panels of Fig. \ref{fig2}. 

\begin{figure}[h!]
\begin{center}
\includegraphics[scale=0.9]{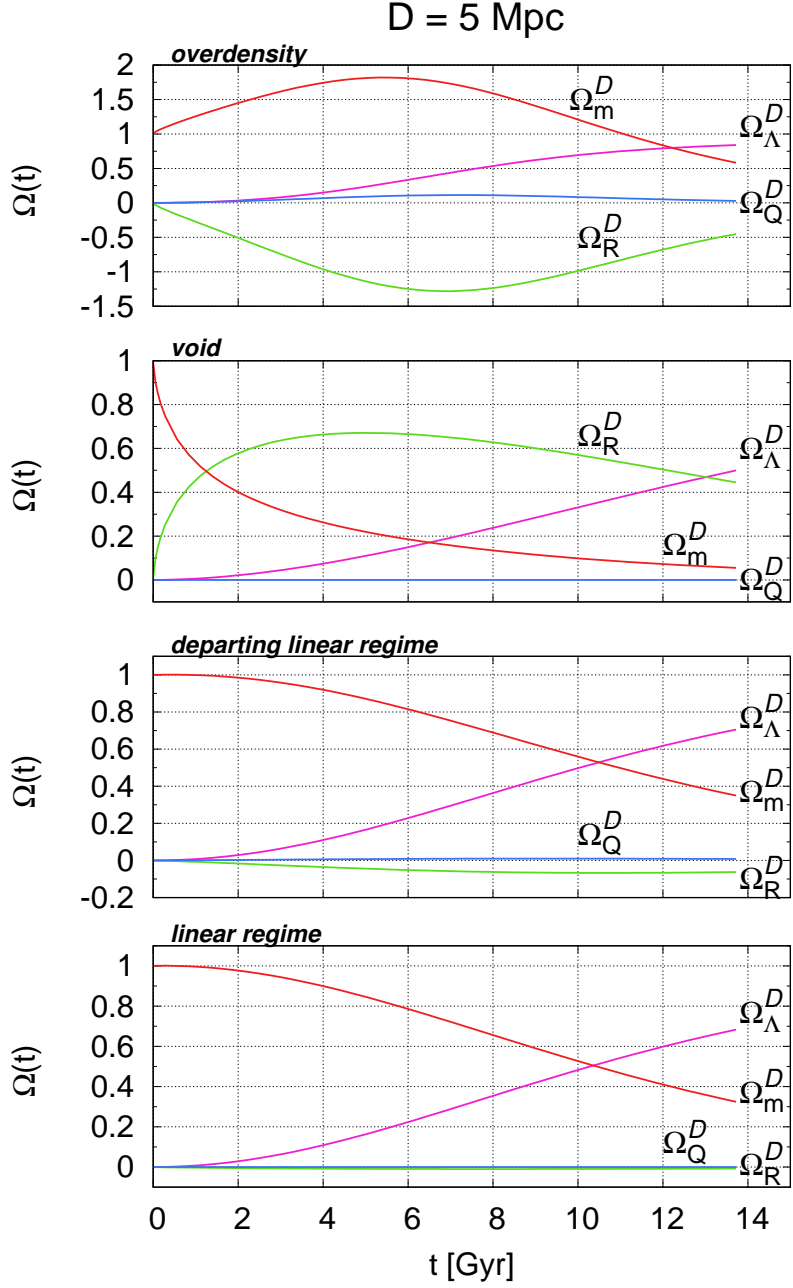}
\end{center}
\caption{Backreaction and evolution of the cosmological system. From top to bottom, these four panels correspond to the places and panels presented in Fig. \ref{fig1}. Each panel presents the evolution of $\Omega$ as defined by eqs. (\ref{cqar}). The kinematic backreaction $\Omega_\mathcal{Q}^{\cal D}$ is depicted with a blue curve. In all cases the kinematic backreaction is rather small: in the linear regime $\Omega_\mathcal{Q}^{\cal D} \sim 10^{-5}$, in the departing linear regime   $\Omega_\mathcal{Q}^{\cal D} \sim10^{-3}$, inside the void $\Omega_\mathcal{Q}^{\cal D} \sim 10^{-7}$, and inside the overdensity  $\Omega_\mathcal{Q}^{\cal D} \sim 0.04-0.07$. The spatial curvature $\Omega_\mathcal{R}^{\cal D}$ is depicted with a green line. Apart from the linear regime, it is non-negligible, and affects the evolution of the cosmological system. The departure from the linear regime (cf. Fig. \ref{fig1}) is characterised by $|\Omega_\mathcal{R}^{\cal D}| > 0.05$.}
\label{fig3}
\end{figure}

By comparing Figs. \ref{fig1} and \ref{fig2} it is apparent that the non-linear evolution is associated with places where the present-day spatial curvature $\Omega_{\cal R}$ is not negligible. It should be pointed out, that at the initial instant the spatial curvature is negligibly small, which means that the non-linear growth is related to the emergence of the spatial curvature. The emergence of the spatial curvature is better depicted in Fig. \ref{fig3}. Figure \ref{fig3} shows the evolution of backreaction and emergence of the spatial curvature at 4 regions presented in Fig. \ref{fig1}. 
The evolution of the {\em cosmic quartet} (\ref{cqar}) at these 4 different locations (presented in Fig. \ref{fig3}) shows that regions which evolve into the non-linear regime have also a substantial build-up of the spatial curvature, and the contribution of the spatial curvature to the evolution of the system is comparable with the contribution from matter. 

There is nothing new or unexpected about this result. The initial perturbations of the spatial curvature, as seen from the Hamiltonian constraint (\ref{hamcon}), are of the same order as density perturbations.
These perturbations are then enhanced in the course of evolution. This also happens in the FLRW regime, where the Friedmannian evolution also allows for the change of the spatial curvature

\[ \Omega_{\cal R}(t) \rightarrow \Omega_k(t) =  -\frac{k}{H^2 a^2}. \]
Thus initial perturbations of curvature are enhanced also by the Friedmannian evolution and its evolution is comparable (but not exactly the same) to the one presented in Fig. \ref{fig3}. This shows that a naive expectation that within a  spatially flat (globally) universe, the spatial curvature is everywhere the same, is simply not accurate. A more realistic expectation is that within the inhomogeneous regions (i.e. locally) $\Omega_{\cal R}$ (or equivalently $\Omega_k$) is of a comparable amplitude as $\Omega_m$.

Another significant result, presented in Fig. \ref{fig3} is that in all cases the contribution from the kinematic backreaction is much smaller than the contribution from the spatial curvature $\Omega_{\cal R}$. This result is consistent with findings reported in Refs. \cite{2006CQGra..23.6379B,2009PhRvD..80l3512W,2013CQGra..30q5006D,2013PhRvD..87l3503B,2013JCAP...10..043R}, which show that even small perturbations of the kinematic backreaction $\Omega_{\cal Q}$ can lead to non-Friedmannian evolution, or as in the case of this Section to a highly non-linear evolution. This shows that the kinematic backreaction ${\cal Q}$ is merely a part of the backreaction phenomenon and non-linear evolution. Thus, care should be exercised when debating the relevance of backreaction using only arguments based on the amplitude of the kinematic backreaction.

The results reported in this Section were obtained based on a model of a single cosmological environment: a pair of a cosmic void and overdensity, with the size of inhomogeneity below 100 Mpc. Beyond that scale the system is homogeneous. The reason for this is that the size of inhomogeneous structures within the Szekeres model increases with radius $r$,  so in order to model realistic structures one has to limit the number of structures to 2 or 3 \cite{2007PhRvD..75d3508B} (but see \cite{2015PhRvD..92h3533S,2016JCAP...03..012S} for an alternative approach). So beyond 100 Mpc in order to eliminate extremely elongated inhomogeneities the model is set to be homogeneous. As a consequence on the scale of 100 Mpc (and beyond) this model is not suitable to study backreaction.

\clearpage

\section{Global evolution and emergence of global spatial curvature within the ensemble of Szekeres models}\label{global}

The model considered in Sec. \ref{local} can only be applied to study inhomogeneities on small scales, i.e. much smaller than the scale of homogeneity (i.e. $\sim 100$ Mpc). In order to study backreaction on a much larger, global scale one needs to implement a different approach to the Szekeres model, such as for example the ensemble approach. In this approach we consider an ensemble of Szekeres worldlines, where each worldline is a separate Szekeres model specified by functions (\ref{szf1})--(\ref{szf5}) with the free parameters set to

\begin{align}
& r_0  = 5 + {\cal U}_{[0-1]} \times 20 {\rm ~Kpc}, \nonumber \\
& \Delta r = \frac{1}{3} r_0, \nonumber \\
& m_0 = -0.0015 + 0.003 \times  {\cal U}_{[0-1]} ,  \nonumber \\
& \eta =  0.5 + 0.3 \times  {\cal U}_{[0-1]}, \nonumber \\
& \rho_i = (1+z_{CMB})^3 \Omega_m 3 H_0^2/(8 \pi G), \nonumber \\
& \Omega_m = 0.315, \nonumber \\
& \Omega_\Lambda = \frac{\Lambda}{3 H_0^2} = 0.685, \nonumber \\
& H_0 = 67.3 {\rm~km} {\rm~s}^{-1} {\rm~Mpc}^{-1}.\label{szes}
\end{align}
\noindent where ${\cal U}_{[0-1]} $ is a random number between 0 and 1 (uniform distribution).

For a large number of realisations, the above prescription provides a wide range of inhomogeneous structures from void-like structures (as in Sec. \ref{local}) to systems with central overdensity and an adjacent void (as for example in Ref. \cite{2007PhRvD..75d3508B}).

Two configurations are investigated:

\begin{enumerate}
\item Swiss-Cheese-type configuration

The Swiss-Cheese configuration has a lattice of a homogeneous FLRW regions and inhomogeneities smoothly approach the FLRW background (as shown in Fig. \ref{fig1}). Through the homogeneous lattice inhomogeneities are joined to other inhomogeneities.
The position of a worldline (with respect to the inhomogeneity defined by (\ref{szes}))
is randomly selected by randomly generating the pseudo-Cartesian coordinates $x,y,z$

\[ x =  2 \, {\cal U}_{[0-1]} r_0, \quad y = 2\,  {\cal U}_{[0-1]} r_0, \quad z = 2 \, {\cal U}_{[0-1]} r_0, \]
which are then used to evaluate the Szekeres coordinates 
\[ r = \sqrt{x^2 + y^2 +z^2}, \quad \theta = {\rm ~arccos} \frac{z}{r}, \quad \phi = {\rm ~arctan} \frac{y}{x}. \]

With a sufficiently large number of worldlines, this approach reproduces a Swiss-Cheese-type configuration, i.e. each type of inhomogeneity with its asymptotically-approaching FLRW region is well mapped. 

\item Styrofoam-type configuration

The Styrofoam-type configuration consists of densely packed closed-cell structures that do not exhibit 
any fixed FLRW lattice. The inhomogeneity is still
defined in the same way as above, i.e. by (\ref{szes}), but the asymptotically-approaching FLRW region is excluded from the Monte Carlo simulation by  selecting only worldlines that are close to the central inhomogeneity, with  coordinates selected as 
\[ r = \sqrt{x^2 + y^2 +z^2}, \quad \theta = {\rm ~arccos} \frac{z}{r}, \quad \phi = {\rm ~arctan} \frac{y}{x}, \]
where 
\[ x = \frac{1}{2}\,  {\cal U}_{[0-1]} r_0, \quad y = \frac{1}{2} \, {\cal U}_{[0-1]} r_0, \quad z =\frac{1}{2}\,   {\cal U}_{[0-1]} r_0.\]

The difference between this type of configuration and the Swiss-Cheese-type configuration is that the inhomogeneous regions are only mapped in their central parts and there is no asymptotically-approaching FLRW region.
On one hand this is an advantage -- no FLRW lattice. On the other hand, this means that inhomogeneities are just stuck together without any attempt to smoothly join them together, which in the Swiss-Cheese configuration is obtained via the FLRW lattice.
\end{enumerate}

The ensemble consists of $10^7$ different and independent worldlines -- since the model is silent ($H_{ab} = 0$, $\nabla_a p =0$, $\pi_{ab} =0$, and $q_a =0$) there is no commutation between the worldlines and therefore each worldline evolves independently. The volume average of a function $A$ within the ensemble of the Szekeres models is 

\begin{equation}
\av{A}_{\cal D} = \frac{ \sum_n A_n v_n} {\sum_n v_n}, 
\end{equation}
where $A_n$ is the value of a function for a specific worldline, and $v_n$ is the volume around each worldline
\begin{equation}
v_n = \sqrt{ | {\rm ~det} \,g| },
\label{volumn}
\end{equation}
  where ${\rm \det} \, g$ is a determinant of the metric (\ref{ds2}). Finally, the size of the domain is 

\[ {\cal D} = V^{1/3} = \left( \sum_n v_n \right)^{1/3}.\]
\noindent For $10^7$ worldlines, the comoving size of the domain of the ensemble defined by (\ref{szes}) is approximately $1050$ Mpc.

\begin{figure*}[h!]
\begin{center}
\includegraphics[scale=0.6]{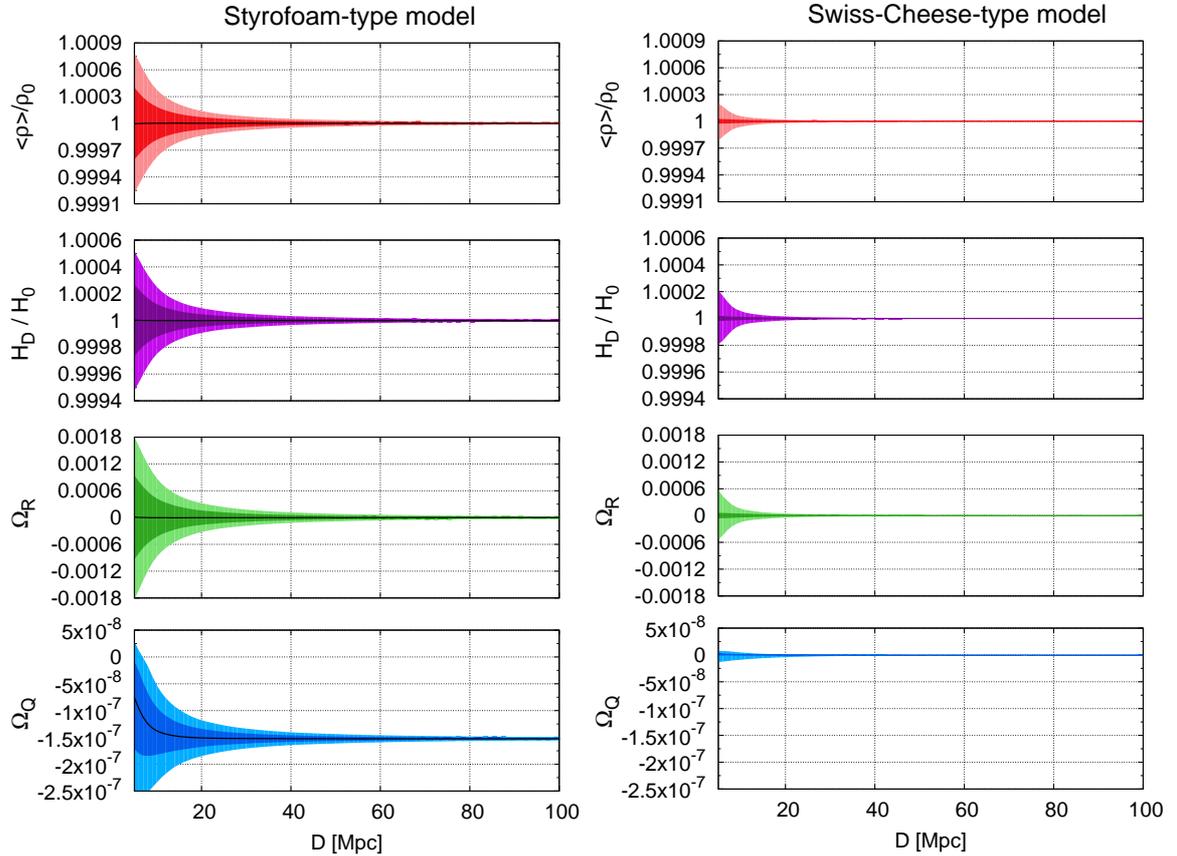}
\end{center}
\caption{Volume average at the initial instant, i.e. at the last scattering instant over a comoving domain of size ${\cal D}$ (at the last scattering instant the physical size of the domain is of order of Kpc not Mpc). Left panels show the results obtained within the Styrofoam-type configuration, and the right panels
within the Swiss-Cheese-type configuration -- for better comparison corresponding right panels have the same scales as the panels on the left.
Upper-most panels show the average density normalised by initial background FLRW density $\rho_i$, second upper panels show the domain Hubble parameter $H_{\cal D}$ normalised by the Hubble parameter at the last scattering $H_i$. Lower panels show
the spatial curvature $\Omega_{\cal R}^{\cal D}$ (second lower), and the kinematic backreaction $\Omega_{\cal Q}^{\cal D}$ (most lower). Darker colour shows the 68\% scatter (i.e. 68\% of values of averaging over random domains of size ${\cal D}$ fall within this interval) and lighter colour shows 95\% (i.e. 95\% of values of averaging over random domains of size ${\cal D}$ fall within this interval). For scales beyond 100 Mpc (comoving) the value of averaging saturates.}
\label{fig4}
\end{figure*}

\begin{figure*}[h]
\begin{center}
\includegraphics[scale=0.6]{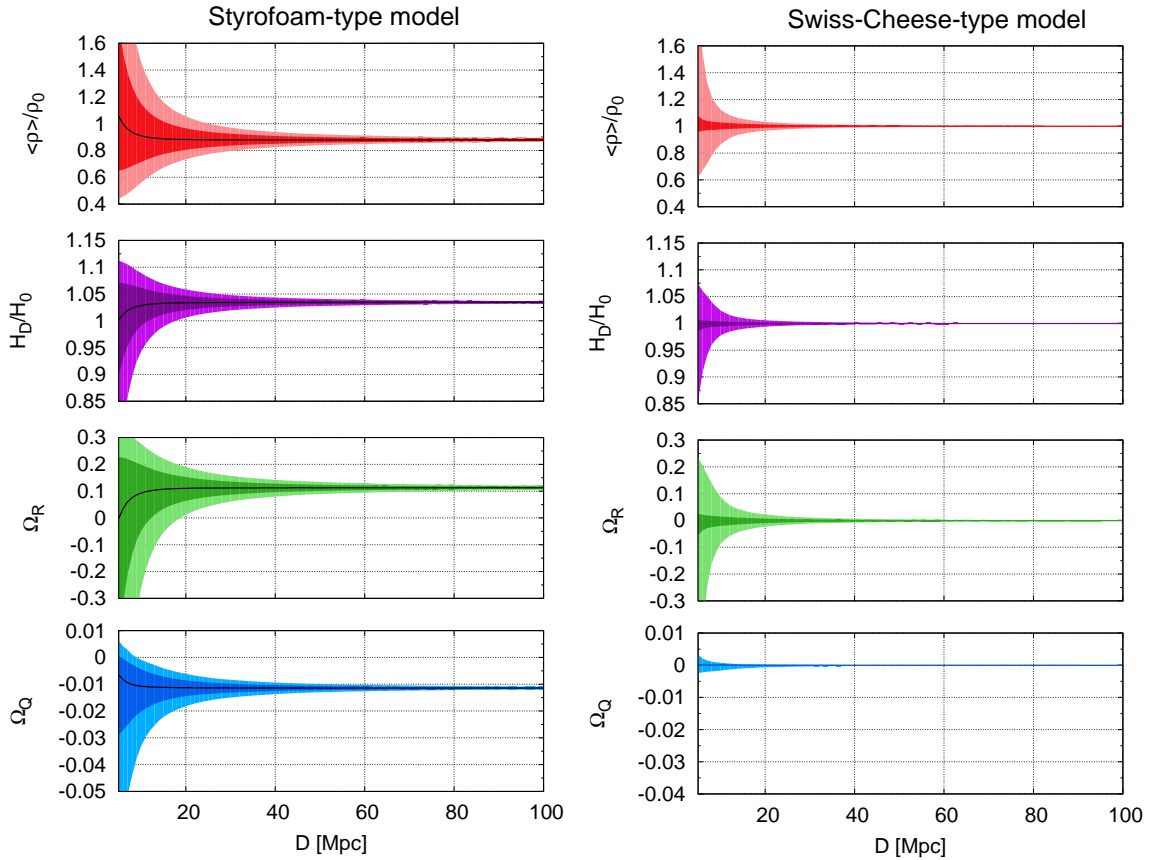}
\end{center}
\caption{Volume averages at the present instant. Left panels show the results obtained within the Styrofoam-type configuration, and the right panels
within the Swiss-Cheese-type configuration -- for better comparison corresponding right panels have the same scales as the panels on the left.
Upper-most panels show the average density normalised by FLRW density $\rho_0$, second upper panels show the domain Hubble parameter $H_{\cal D}$ normalised by the Hubble constant $H_0$. Lower panels show
the spatial curvature $\Omega_{\cal R}^{\cal D}$ (second lower), and the kinematic backreaction $\Omega_{\cal Q}^{\cal D}$ (most lower). Darker colour shows the 68\% scatter (i.e. 68\% of values of averaging over random domains of size ${\cal D}$ fall within this interval) and lighter colour shows 95\% (i.e. 95\% of values of averaging over random domains of size ${\cal D}$ fall within this interval). For scales beyond 100 Mpc the cosmic variance is negligible.}
\label{fig5}
\end{figure*}

As in Sec. \ref{local}, the initial instant is set to be the last scattering instant. The state of the ensemble at the initial instant presented in Fig. \ref{fig4}, shows that initially the model is quite homogeneous,
with average density $\av{\rho}$ and domain expansion rate $H_{\cal D}$ quickly approaching the $\Lambda$CDM values, and the spatial curvature and kinematic backreaction approaching $\, 0 \,$ when averaged over a sufficiently large domain. As seen from Fig. \ref{fig5}, after 13.8 Gyr of evolution the statistical homogeneity is still present at scales beyond 100 Mpc --- the averaging over random domains of size ${\cal D} > 100 {\rm~Mpc}$ produces the same results (i.e. the cosmic variance is negligible small at these scales). However, the mean values depend on the global model, and the results differ between the Swiss-Cheese-type and Styrofoam-type models.

The mean evolution within the Swiss-Cheese-type model follows the background $\Lambda$CDM model: average density $\av{\rho} = \rho_0$, domain expansion parameter $H_{\cal D} = H_0$, the model is practically spatially flat with  $\Omega_{\cal R}= 2.4 \times 10^{-4}$, and the kinematic backreaction is negligibly small $\Omega_{\cal Q}= -1.6 \times 10^{-5}$.
However, within the Styrofoam-type model the average density falls below the $\Lambda$CDM, i.e.  $\av{\rho} = 0.88 \, \rho_0$ and the domain expansion parameter is faster than in $\Lambda$CDM background $H_{\cal D} = 1.03 \, H_0$. In addition, the mean spatial curvature is $\Omega_{\cal R}= 0.11$, and the kinematic backreaction is $\Omega_{\cal Q}= -1.1 \times 10^{-2}$, which shows significant deviation from the Friedmannian evolution of the background model.

\clearpage

\begin{figure*}[h]
\begin{center}
\includegraphics[scale=0.65]{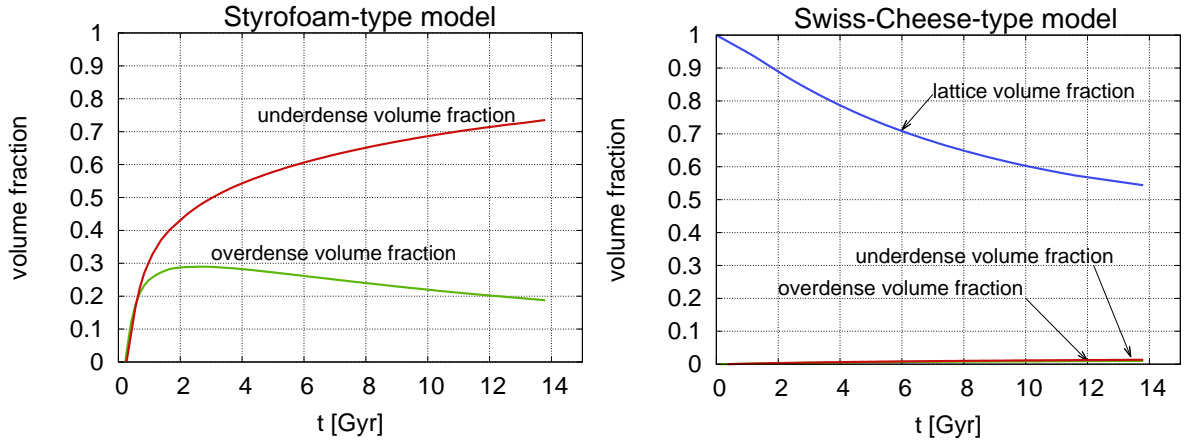}
\end{center}
\caption{The volume fraction occupied be underdense, overdense, and $\Lambda$CDM-type regions within a domain of ${\cal D} \approx 1050$ Mpc. Within the Styrofoam-type model (left panel) the total volume is dominated by underdense regions, which at the present-day occupy approximately $75\%$ of the total volume. Within the Swiss-Cheese-type model (right panel) regions  that only deviate by less than $1\%$ from the $\Lambda$CDM model occupy more than $50\%$ of the total volume; the contribution from the underdense ($\rho<0.9 \, \rho_{\Lambda CDM}$) and overdense ($\rho>1.1 \,\rho_{\Lambda CDM}$) is  merely at the percent level, and the rest of the volume is occupied by almost $\Lambda$CDM-like regions.}
\label{fig6}
\end{figure*}

The explanation for the difference between the global behaviour of the Swiss-Cheese-type model and the Styrofoam-type model is presented in Fig. \ref{fig6}.
Figure \ref{fig6} shows the volume fraction of various types of structures within the Styrofoam-type and Swiss-Cheese-type models.
The underdense fraction is defined as
\[ f_u = \frac{ V_{underdense} } {V_{total} }, \]
where $V_{underdense}$ is volume occupied by regions with $\rho<0.9 \, \rho_{\Lambda CDM}$.
The overdense fraction is defined as
\[ f_o = \frac{ V_{overdense} } {V_{total} }, \]
where $V_{overdense}$ is volume occupied by regions with $\rho>1.1 \, \rho_{\Lambda CDM}$.
Finally, the lattice fraction is defined as
\[ f_l = \frac{ V_{lattice} } {V_{total} }, \]
where $V_{lattice}$ is volume occupied by regions with $0.99 \, \rho_{\Lambda CDM} < \rho < 1.01 \, \rho_{\Lambda CDM}$.

As seen the Swiss-Cheese model is dominated by the asymptotically-homogeneous regions  that only deviate by less than $1\%$ from the $\Lambda$CDM model. Such regions occupy more than $50\%$ of the total volume of the Swiss-Cheese-type model, while the contribution from the underdense ($\rho<0.9 \, \rho_{\Lambda CDM}$)
and overdense ($\rho>1.1 \,\rho_{\Lambda CDM}$) is  merely at the percent level, and the rest of the volume is occupied by almost $\Lambda$CDM-like regions.
Consequently, the averages and the mean global evolution follows closely the $\Lambda$CDM model.
Contrary to the Swiss-Cheese model, within the ensemble of the Szekeres models of the Styrofoam type, 
the Monte Carlo simulation only probes the central parts of inhomogeneity and avoids the asymptotic FLRW regions. 
As a result, the volume is dominated by underdense regions with $f_u = 0.75$ at the present-day instant, and the overdense regions occupy $20\%$ of the total volume.
The reason why underdense regions dominate the volume is simply because voids expand faster than the overdense regions, and so they quickly 
start to occupy larger factions of the volume. Interestingly, by comparing the time scales of Fig. \ref{fig6} to Fig. \ref{fig1} one can see that the instant when voids start to dominate the total  volume is when they enter the non-linear regime.

\begin{figure*}
\begin{center}
\includegraphics[scale=0.65]{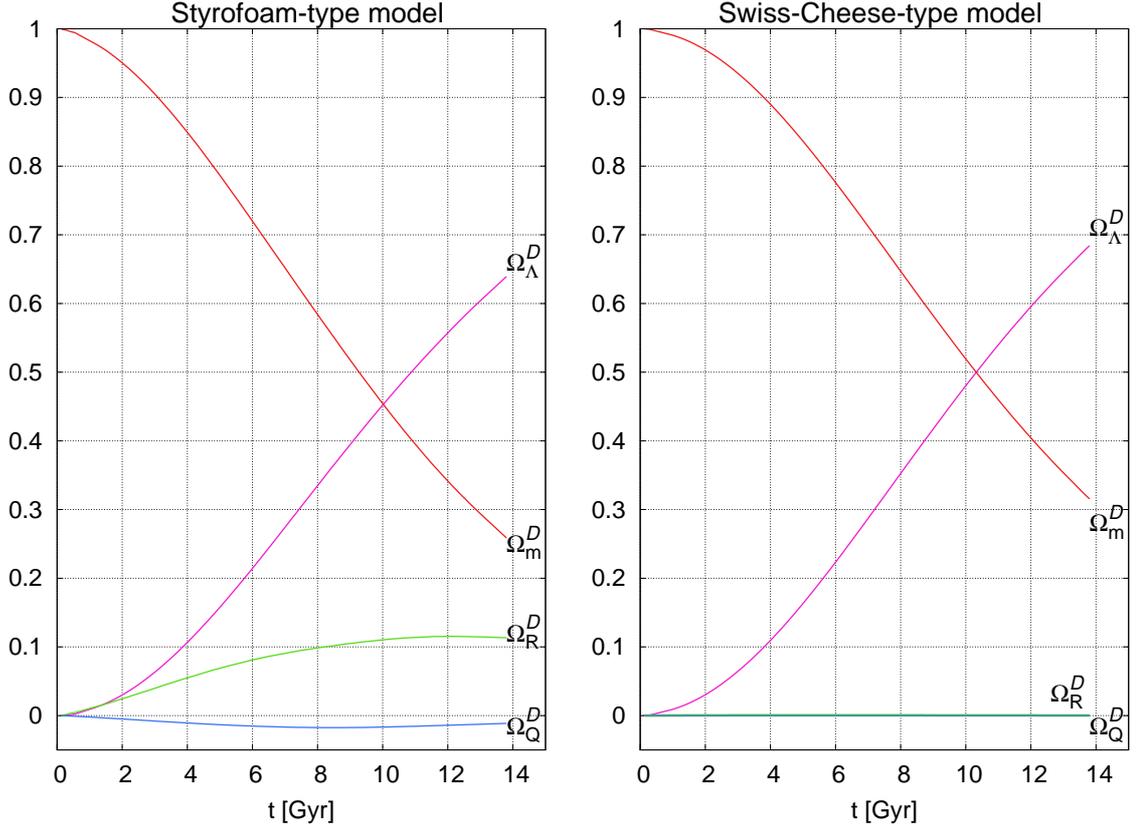}
\end{center}
\caption{The evolution of the {\em cosmic quartet} (\ref{cqar}) within the ensemble of Szekeres models evaluated within a comoving domain of size 1050 Mpc. Since the cosmic variance is negligible at these scales (cf. Figs. \ref{fig4} and \ref{fig5}) this plot shows the evolution on the mean matter density (red), dark energy (magenta), spatial curvature (green), and kinematic backreaction (blue). }
\label{fig7}
\end{figure*}

The evolution of the {\em cosmic quartet} (\ref{cqar}) is presented in Fig. \ref{fig7}. Not surprisingly, since the volume of the Swiss-Cheese-type model is dominated by the $\Lambda$CDM-like regions, the evolution of the {\em cosmic quartet} follows the $\Lambda$CDM behaviour. As for the Styrofoam-type model, which volume is dominated by voids, the evolution of the {\em cosmic quartet}, when compared with Fig. \ref{fig3}, resembles the void-like behaviour.
It is worth pointing out that within the Styrofoam-type model there is no fixed background.
The mean density evolves slightly differently than in the $\Lambda$CDM model, which was used to specify the model at the initial instant. In particular, the evolution of the mean spatial curvature is unlike in the $\Lambda$CDM  which is spatially flat.

The emergence of the mean global spatial curvature (within the Styrofoam-type model) has an intuitive explanation. Within this model underdense regions dominate from the start, but since density contrast of overdense regions increase at a slightly higher rate -- this is because, as seen from eq. (\ref{ffe1}) both density and shear negatively contribute to $\dot{\Theta}$ -- thus overdense regions more quickly pass the $10\%$-threshold used in the definition of $f_o$ and $f_u$, and thus in the left panel of Fig. \ref{fig6} it looks like overdense regions slightly dominate in the first 500 My of evolution.
At that instant the spatial curvature of the Styrofoam-type model is $\Omega_{\cal R} \approx 5\times 10^{-3}$.  After that instant $f_u > f_o$ and soon afterwards the amplitude of $\Omega_{\cal R}$ becomes non-negligible.
This phenomenon can also be understood in terms of the analysis presented in Ref. \cite{2011CQGra..28p5004R}, which focused on the dynamical system of eqs. (\ref{bucherteq1})--(\ref{bucherteq3}). The analysis of the dynamical system (\ref{bucherteq1})--(\ref{bucherteq3}) showed that even a tiny perturbation in ${\cal Q}_{\cal D}$ can drive the system into another basin of attraction that is dominated by the averaged
curvature \cite{2011CQGra..28p5004R}. This property of a global gravitational instability of the Friedmannian model has been identified as the reason for large curvature deviations. The Styrofoam-type model based on the ensemble of the Szekeres models provides an explicit realisation of this scenario.

\section{Conclusions}\label{conclusions}

The models considered in this paper (local environment and global ensemble) addressed the issue of the amplitude of backreaction, and its impact on the evolution  of a cosmological system. 
The analysis was based on the exact, cosmological, and non-symmetrical solutions of the Einstein equations, i.e. the Szekeres model, and it provided examples of relativistic models with non-vanishing backreaction (cf. \cite{Buchert:2015iva}). 
The obtained results show that the non-linear evolution and backreaction are closely associated with the spatial curvature. The growth of inhomogeneities cannot be separated from the growth of the spatial curvature, which variation across the Universe is comparable with the variation in the matter field, and in addition the global mean spatial curvature does not necessarily average out to zero. This implies that the emergence of the Cosmic Web in the real Universe should also be associated with the emergence of the mean spatial curvature.  

There is nothing ``exotic'' about the obtained results, and the second-order effects do not ``magically'' appear with 
an amplitude a few orders of magnitude larger than the first order effects. These findings can easily be understood in a simple and logical manner.
Initially, at the last scattering instant, the Universe is fairly homogeneous with only small perturbations present.  As long as the growth of structures is linear and perturbations are small, the differences in the expansion rates are negligible and both types of regions: underdense and overdense expand at a similar, background rate.
Once the growth of cosmic structure is non-linear (e.g. matter shear $\sigma^2$ is no longer negligible) the expansion rate of overdense regions efficiently slows down (cf. eq. (\ref{ffe1})).
When that happens  the volume of the Universe becomes dominated by voids (cf. Fig. \ref{fig6}). Once the volume of the Universe starts to be dominated by voids then the total volume (eg. within the cosmic horizon) increases faster than in the FLRW model, and since matter is conserved, the mean density is lower than in the  $\Lambda$CDM model.
This explains why, as seen in Fig. \ref{fig5}, the mean density is below the $\Lambda$CDM matter density, and the expansion rate is slightly higher than $H_0$. 
As for the spatial curvature, the density perturbations are coupled with expansion rate perturbations and curvature perturbations (cf. eq. \ref{hamcon})). Thus, initially at the last scattering instant, not only density but also tiny curvature perturbations are present. In the course of evolution, these perturbations also grow (in the FLRW regime, $\Omega_{\cal R} \to \Omega_k = - k/\dot{a}^2$), which leads to large variations of spatial curvature across cosmic structures (cf. Fig. \ref{fig2}). In addition, since the volume of the Universe in the non-linear regime is dominated by voids, the mean global spatial curvature evolves from $\Omega_{\cal R} =  0$ to $\Omega_{\cal R} \approx 0.1$ (cf. Fig. \ref{fig7}).

These findings were obtained based on the Styrofoam-type model. In the Swiss-Cheese model, at the local scales the picture is similar (cf. Figs. \ref{fig1}--\ref{fig3}) --- i.e. non-linear evolution leads to large differences between expansion rates of underdense and overdense regions, as well as, large variations of the spatial curvature --- however on global scales the results are different, which can be linked to a fact that in the Swiss-Cheese model the volume is dominated by the $\Lambda$CDM-like regions. 
Since the volume of the real Universe is in fact dominated by cosmic voids \cite{2012MNRAS.421..926P}, it is reasonable to conclude that the Styrofoam-type model is more realistic than the Swiss-Cheese model.
As a results, one can expect that the global mean spatial curvature of our Universe should also be dominated by voids, and subsequently it should deviate from zero in the low-redshift Universe,
leading to its evolution from spatial flatness ($\Omega_{\cal R} = \Omega_k = 0$) in the early Universe to a negative spatial curvature ($\Omega_{\cal R} \sim \Omega_k > 0$) at the present day epoch, in a similar manner as presented in Fig. \ref{fig7}.

It is interesting to point out that, in fact, the analysis of low-redshift data, such as a supernova data alone (without combining it with the CMB) implies large spatial curvature, i.e. $\Omega_k \approx 0.2$ and only after inclusion of the CMB reduces to $\Omega_k  = 0.005 \pm 0.009$  \cite{2014A&A...568A..22B}.
Also, supernova data alone point towards slightly lower values of $\Omega_m$ compared to the CMB constraints, which could be understood, not just in terms of various systematics, but also partly in terms of findings presented in Fig. \ref{fig7}.
In addition, there is a known tension between the values of $H_0$ derived from the CMB \cite{2016A&A...594A..13P} and the local measurements \cite{2016ApJ...826...56R}
which again, apart from various systematics \cite{2016arXiv160802487P}, could also be partly explained in terms of findings presented in Fig. \ref{fig5}.

The models like the one presented in this paper, i.e. Styrofoam-type model,
are not perfect realisations of our Universe. There are a number of limitations, for example the lack of rotation $w_{ab}$ excludes presence of virialised regions; the lack of magnetic Weyl tensor $H_{ab}$ excludes multiple eigenvalues of shear; and lack of heat flow $q_a$ excludes energy transfers from one cosmic cell to another. This all means that one should exercise caution when drawing conclusion as to the properties of our Universe. Still, the obtained results should encourage further studies of the cosmological backreaction and 
development of numerical cosmology towards realistic models of the Universe
\cite{Bentivegna:2015flc,Mertens:2015ttp,2017PhRvD..95f4028M}. Also, these results should encourage cosmologists to think outside the box, especially when dealing with low-redshift data, and allow for example for $\Omega_k \ne 0$, even if at the CMB the Universe was spatially flat\footnote{Separation of the low-$z$ from high-$z$ data has already been implemented by some cosmologists, for example in Ref. \cite{2007PThPh.117.1067K} the supernova analysis was conducted using different sets of cosmological parameters for low-$z$ and high-$z$ data; in Ref. \cite{2010JCAP...08..023V} the CMB data was analysed independently from the low-$z$ cosmology that enters via the distance to the last scattering surface.}.
Allowing for $\Omega_k$ to be a free parameter at low-$z$ and independent from high-$z$ constraints, could in principle reduce some of the  tensions and inconsistencies in the observed data \cite{2016IJMPD..2530007B,2016arXiv161201529C}, but the actual analysis remains to be done.
In addition, cosmologists should also aim at 
directly measuring the curvature of the low-redshift Universe to check if it indeed deviates from the CMB constraints \cite{2015PhRvL.115j1301R}.

\acknowledgments
I would like to thank Thomas Buchert, Jan Ostrowski, Boudewijn Roukema, David Wiltshire, and anonymous referee for their comments and suggestions. This work was supported by the Australian Research Council through the Future Fellowship FT140101270. 

\bibliography{backszek-rev} 
\bibliographystyle{ieeetr_arXiv}

\end{document}